\documentclass[aps,prl,twocolumn,groupedaddress,amsmath,amssymb,showpacs,letterpaper]{revtex4}
\usepackage{graphicx} 
\usepackage{epstopdf} 
\usepackage{dcolumn} 
\usepackage{bm} 
\usepackage{relsize}

\begin{document}

\title{Controlling the spin orientation of photoexcited electrons by symmetry breaking}
\author{Lan~Qing}
\email{lan.qing@rochester.edu}
\author{Yang~Song}
\author{Hanan~Dery}\altaffiliation[Also at ]{Department of Electrical and Computer Engineering, University of Rochester, Rochester, NY 14627.}
\affiliation{Department of Physics and Astronomy, University of Rochester, Rochester, New~York 14627}
\begin{abstract}
We study reflection of optically spin-oriented hot electrons as a means to probe the semiconductor crystal symmetry and its intimate relation with the spin-orbit coupling. The symmetry breaking by reflection manifests itself by tipping the net-spin vector of the photoexcited electrons out of the light propagation direction. The tipping angle and the pointing direction of the net-spin vector are set by the crystal-induced spin precession, momentum alignment and spin-momentum correlation of the initial photoexcited electron population. We examine non-magnetic semiconductor heterostructures and semiconductor/ferromagnet systems and show the unique signatures of these effects.
\end{abstract}
\pacs{72.25.Fe, 72.25.Mk, 72.25.Rb, 71.70.Ej}
\maketitle

The intimate relation between the spin-orbit coupling of crystals and their symmetry is a theme of on-going research for more than a half-century \cite{Dresselhaus,Luttinger_Kohn,Dyakonov_Perel,Dymnikov_JETP76,Meier_OO,Pengke_PRL10,Zemskii_JETP76,Alperovich_JETP80,Bychkov-Rashba,Bhat_PRL00,Wundelich,Crooker,Kanamori_63,Dietl_PRB01}. The rapid progress of spintronics research has provided new powerful techniques to study this relation \cite{Zutic}. In semiconductors it is readily seen in the valence band energy dispersion \cite{Dresselhaus,Luttinger_Kohn}, in spin relaxation of electrons \cite{Dyakonov_Perel}, or in optical selection rules \cite{Dymnikov_JETP76,Meier_OO}. In magnetic materials this relation sets the magnetocrystalline anisotropy and magnetostriction constants \cite{Kanamori_63,Dietl_PRB01}. In this letter, we study intriguing results of this relation which depend on the spin-momentum correlation and coherent spin precession of photoexcited electrons in bulk semiconductors.

Using symmetry breaking by reflection we show that the net-spin vector of photoexcited electrons is tipped away from the light propagation axis. We explain the dependence of the tipping on spin-momentum correlation and coherent spin precession, and then calculate the tipping angles with partial, complete, nonmagnetic and magnetic reflections. In prospective experiments, one can measure signatures of the correlation and coherent precession by probing the tipped net-spin vector rather than the coherent distribution of photoexcited electrons.  One clear advantage is that the net-spin vector decays orders of magnitude slower than the coherence time.

\begin{figure}
   \centering
   \includegraphics[width=8.6cm]{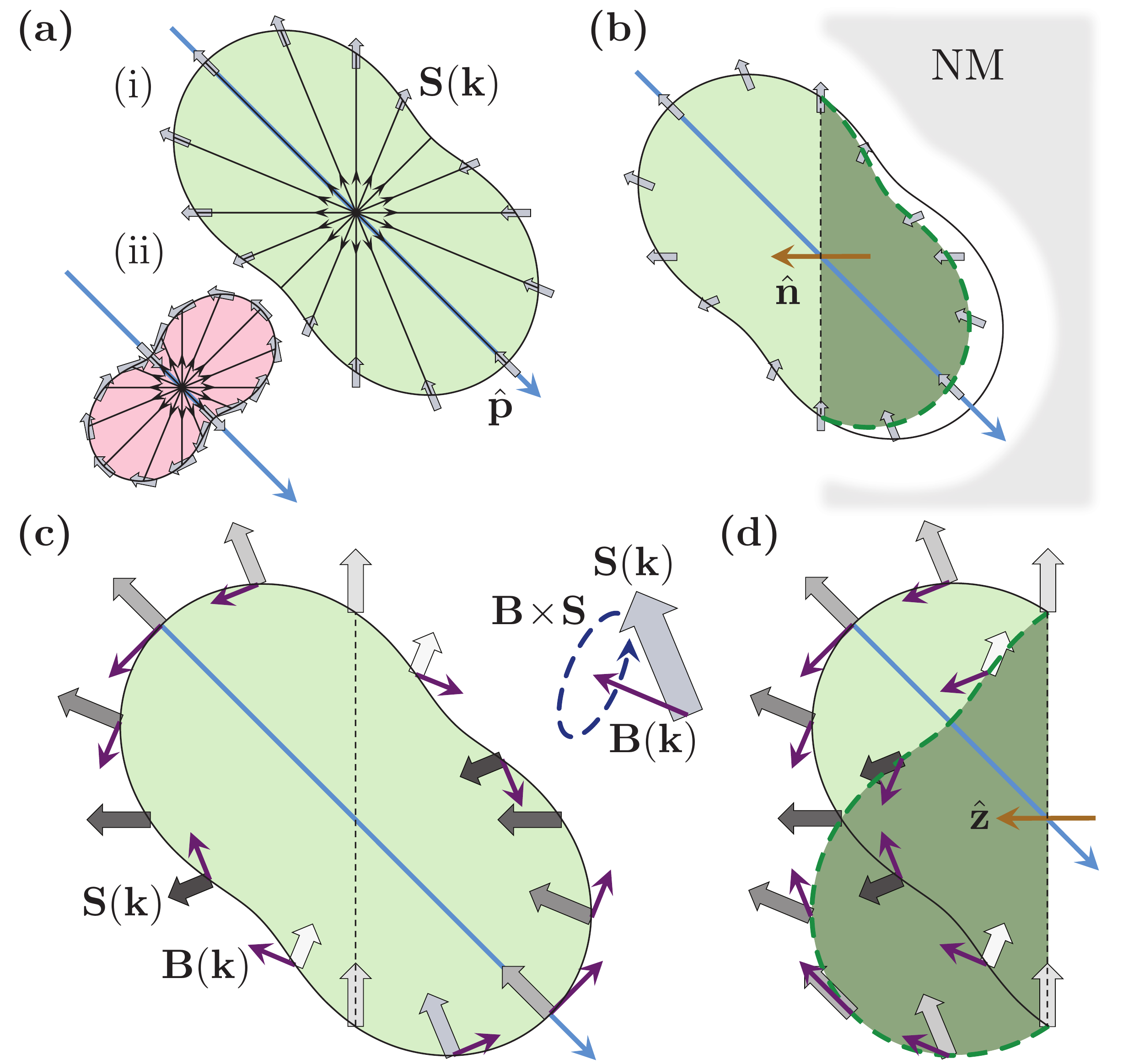}
   \caption{(a) Momentum and spin distributions of photoexcited electrons following transitions with (i) heavy holes  and (ii) light holes (excitation with heavy holes is stronger due to their larger density of states). The length of a line from the center represents the relative population of excited electrons with momentum along the arrow's direction. The thick arrows on the edge line represent the correlated spins. (b) Correlation induced tipping by partial reflection.  The net-spin vector is tipped away form $\hat{\mathbf{p}}$ due to missing transmitted electrons with spins mostly along the interface normal $\hat{\mathbf{n}}$. (c) Spin precession of electrons immediately after photoexcitation. Electrons moving in opposite directions have the same initial spin direction but precess at opposite angles. (d) Precession induced tipping by complete reflection. The net-spin vector is tipped away from $\hat{\mathbf{p}}$ due to rephasing (see text).}
   \label{fig:scheme}
\end{figure}

We first explain the correlation tipping phenomenon in a photoexcited direct gap semiconductor. Figure~\ref{fig:scheme}(a) shows the momentum alignment and spin-momentum correlation of electrons immediately after photoexcitation. The shown distributions are compactly derived by using the spherical model at the top of the valence band \cite{Luttinger_Kohn} and $s$-type states in the conduction band. The density matrix of photoexcited electrons is then \cite{Dymnikov_JETP76,Meier_OO},
\begin{eqnarray}
\!\!\!\!\!\!\!\!\!\!\! \mathcal{D}(t,\mathbf{k})\!\!&=&\!\!\Big\lbrace \mathcal{I}+\tfrac{1}{2}\alpha(t)\big[3\big|\hat{\mathbf{e}}\cdot\hat{\mathbf{k}}\big|^2\!\!-\!1\big]\mathcal{I} +2S(t)(\hat{\bm{\sigma}}\cdot\hat{\mathbf{p}}) \nonumber  \\&
+&\tfrac{1}{2}\beta(t)\big[3(\hat{\bm{\sigma}}\cdot\hat{\mathbf{k}})(\hat{\mathbf{p}}\cdot\hat{\mathbf{k}})-(\hat{\bm{\sigma}}\cdot\hat{\mathbf{p}})\big]\Big\rbrace F(t,k)\,.  \label{eq:dymnikov}
\end{eqnarray}
$\hat{\mathbf{k}}$ ($\hat{\mathbf{e}}$) is the unit vector in the direction of electron momentum (light polarization). $\hat{\bm{\sigma}}$ ($\mathcal{I}$) denotes the Pauli matrix vector (the $2\times2$ unit matrix). The photon angular momentum unit vector is defined by $\hat{\mathbf{p}} \equiv i\,\hat{\mathbf{e}}\times\hat{\mathbf{e}}^*$. For linearly polarized light $\hat{\mathbf{p}}=0$ and for circularly polarized light, $\big|\hat{\mathbf{e}}\cdot\hat{\mathbf{k}}\big|^2=[1-(\hat{\mathbf{p}}\cdot\hat{\mathbf{k}})^2]/2$. The parameters $\alpha$, $S$ and $\beta$ are, respectively, measures of momentum alignment,  of average spin and of spin-momentum correlation. $S(t)$ decays exponentially with the spin relaxation time from initial value $S_0\approx-1/4$. $\alpha(t)$ and $\beta(t)$ decay exponentially with the much shorter momentum relaxation time from their initial values $\alpha_0\approx\beta_0\approx \pm 1$ \cite{footnote1}. The lower/upper sign is for transitions with heavy/light holes. Finally, $F(t,k)$ relates to the density of excited electrons. Figure~\ref{fig:scheme}(b) elucidates the correlation induced tipping effect for electrons that are generated with heavy holes. This effect is governed by transmission of electrons immediately after excitation (prior to the decay of $\beta(t)$ in Eq.~(\ref{eq:dymnikov})). For example, consider electrons that propagate parallel and perpendicular to the reflection plane (shown in Fig.~\ref{fig:scheme}(b) by the two spins on the edges of the dash line and the two spins along $\pm\hat{\mathbf{n}}$). Their net spin is not collinear with $\hat{\mathbf{p}}$ due to missing electrons along $-\hat{\mathbf{n}}$. Similar explanation holds for other spin and momentum directions. The overall effect is that the net-spin vector changes from $S_0\hat{\mathbf{p}}$ to $S_0\left[(1-\delta_a)\hat{\mathbf{p}}-\delta_b(\hat{\mathbf{p}}\cdot\hat{\mathbf{n}})\hat{\mathbf{n}}\right]$ where $\delta_a$ and $\delta_b$ are measures of the transmission amplitude.

The second tipping phenomenon results from intrinsic spin precession of electrons in semiconductors without inversion symmetry \cite{Dyakonov_Perel}. The precession is due to a torque exerted by the electron's effective magnetic field whose components are $B_i \propto k_i(k_j^2-k_m^2)$ where $\{k_i,k_j,k_m\}$ are the electron's wavevector components along the crystallographic axes. Figure~\ref{fig:scheme}(c) shows the intrinsic spin precession of photoexcited electrons immediately after generation. Electrons that move in opposite directions have similar initial spin direction but precess at opposite angles. The net angle precession of the pair is zero on average. This picture is changed by specular reflection of one of the electrons off an interface as shown in Fig.~\ref{fig:scheme}(d). Here, $k_z$ and $B_z$ of an electron flip signs after specular reflection whereas its spin components $\{S_x, S_y, S_z\}$ remain unchanged. This causes $S_x$ and $S_y$ to `re-phase' to some degree due to the flip of $B_z$ whereas $S_z$ keeps its precession unperturbedly. The net effect is that transverse spin components with respect to the normal of the reflection plane decay slower than the longitudinal component. This effect is robust in both complete or partial reflection and it can be further amplified by multiple interface reflections. Scattering events reduce the magnitude of the net-spin vector but they do not change its direction. We will show that in spite of a notable Dyakonov-Perel spin relaxation of hot electrons the net-spin vector after energy relaxation remains measurable.

The time evolution of photoexcited electrons is studied by extensive Monte Carlo simulations \cite{auxiliary}. The initial wavevector and spin directions are randomized according to the distributions in Eq.~(\ref{eq:dymnikov}). The initial electron density follows the light attenuation profile. We use the effective mass approximation in calculating electron transport and quantum mechanical transmission probabilities across interfaces. Effective masses, band-gaps and band offsets are taken from Ref.~\cite{Adachi_PAGA}. Momentum and energy relaxations of hot electrons are simulated by the Fr\"{o}hlich interaction \cite{Cardona_Book_ch5}. Between scattering and reflection events the spin precesses about its intrinsic field. The simulation ends when electrons reach the bottom of the conduction band (typically after 1~ps). Spin relaxation at later times occurs on $>$1~ns time scales \cite{Crooker}.

\begin{figure}
   \centering
   \includegraphics[width=8.6cm]{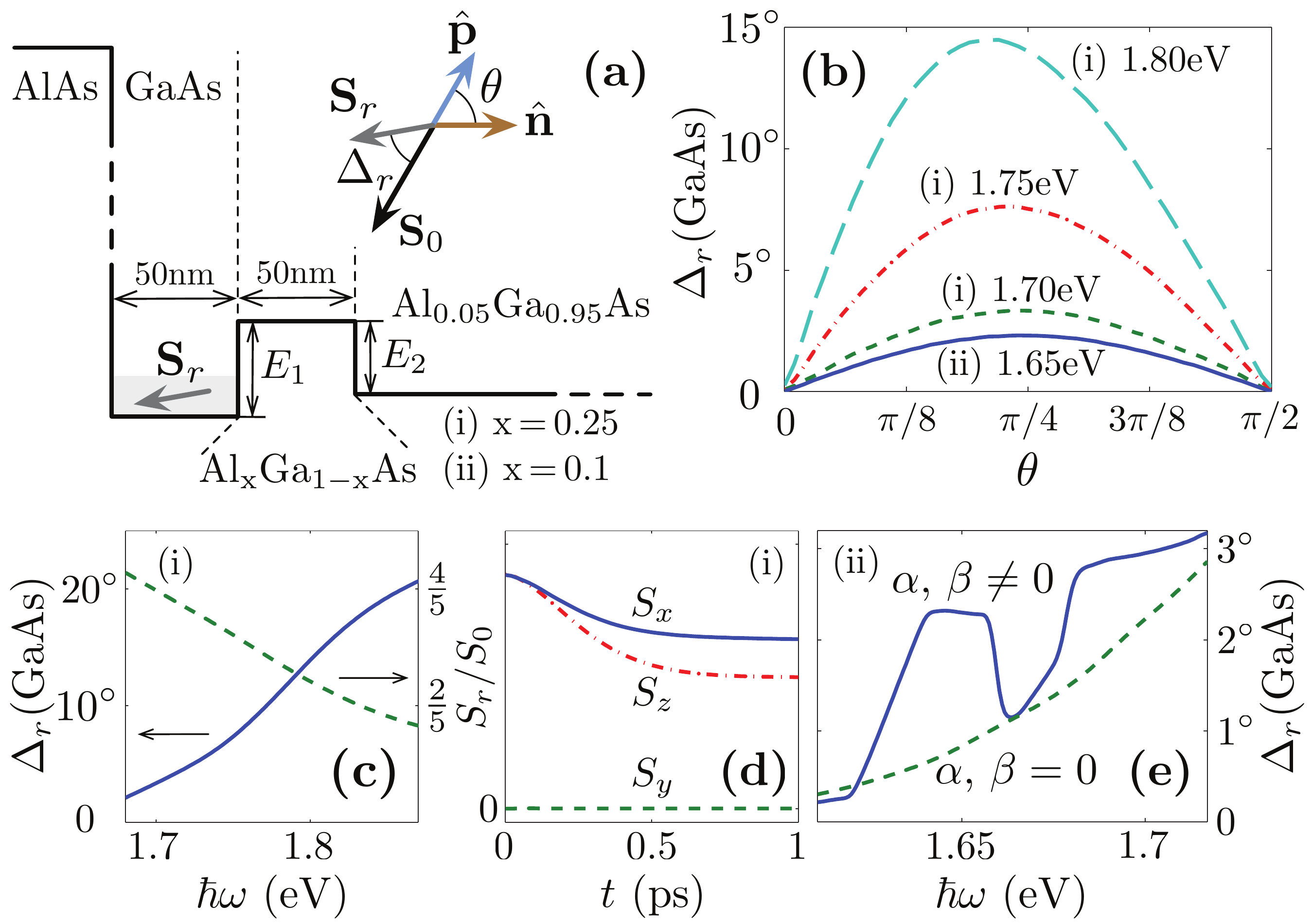}
   \caption{(a) Simulated excitation geometry and heterostructure setups (i) and (ii). $\hat{\mathbf{n}}$ is along the $z$ crystal axis. $\hat{\mathbf{p}}$ is in the $x$-$z$ crystal plane ($\cos\theta=\hat{\mathbf{p}}\cdot\hat{\mathbf{n}}$). $\mathbf{S}_r$ is the remained net-spin vector in the GaAs region after energy relaxation. $\Delta_r$ is the tipping angle of $\mathbf{S}_r$ away from the optically injected spin direction. (b) $\Delta_r$ vs $\theta$ for various setups and photon energies.  (c) $\Delta_r$ (left) and $S_r$ (right) vs photon energy in setup (i). The decay of $\mathbf{S}_r$ is due to hot-electron spin relaxation. (d) Spin evolution along crystal axes in the GaAs region in setup (i). (e) $\Delta_r$ vs photon energy in setup (ii) with/without alignment and correlation. (c)-(e) correspond to $\theta=\pi/4$.}
   \label{fig:Al}
\end{figure}

We first study the tipping effects in non-magnetic heterostructures. To distinguish between correlation and precession induced tipping we employ two setups of the heterostructure $\rm{AlAs/GaAs(50\,nm)/Al_{x}Ga_{1-x}As(50\,nm)}$ $\rm{/Al_{0.05}Ga_{0.95}As}$ \cite{footnote_Al5Ga95}. The structure, shown in Fig.~\ref{fig:Al}(a), ensures that confinement effects are negligible and all regions are bulk in nature. Setup (i) includes a high inner barrier (x=0.25) for which the potential steps are $E_1=230$~meV \& $E_2=185$~meV (see Fig.~\ref{fig:Al}(a)). The tipping angle is governed by spin precession of hot electrons in the GaAs region prior to thermalization while bouncing back and forth from the potential walls. Setup (ii) includes a shallower inner barrier (x=0.1) for which the potential steps are $E_1=90$~meV \& $E_2=45$~meV. For certain photon energies there is  a favored net transmission from the GaAs region to the $\rm{Al_{0.05}Ga_{0.95}As}$ region. The spins of transmitted electrons are mostly aligned with the interface normal rather than the optically injected spin direction. In both setups, quantum tunneling across the 50~nm inner barrier is negligible. 

Figure~\ref{fig:Al}(b) shows the dependence of the tipping angle on the light propagation direction for various photon energies. $\Delta_r$ is the tipping angle of the net-spin vector in the GaAs region after energy relaxation. $\theta$ is the angle between the photon angular momentum ($\hat{\mathbf{p}}$) and the interface normal ($\hat{\mathbf{n}}$) as shown in Fig.~\ref{fig:Al}(a). We first focus on the results of setup (i). The tipping angle is maximized when light propagates along the $\langle 101\rangle$ \& $\langle 011\rangle$ crystallographic axes ($\theta=\pi/4$) due to the faster spin precession along these directions. Similarly, the tipping angles increase with photon energies due to the enhanced precession of electrons. Figure~\ref{fig:Al}(c) shows the tipping angle as a function of photon energy when $\theta=\pi/4$. The net-spin vector after energy relaxation remains measurable in all of the studied excitation range. Figure~\ref{fig:Al}(d) shows the time evolution of the net-spin along the crystallographic axes for $\theta=\pi/4$ and photon energy of 1.75~eV. The aforementioned rephasing effect is seen in the slower decay of $S_x$ during the energy relaxation in the first 0.5~ps (the $z$ component of the intrinsic field changes direction with each reflection). At later times ($>$1~ps) the precession frequency and spin relaxation are much slower.

Figure~\ref{fig:Al}(e) shows correlation induced tipping angles as a function of photon energy when $\theta=\pi/4$. As a reference, we also simulate a case where momentum alignment and spin-momentum correlation are neglected (dash line). In the latter case, only spin precession during the initial coherent phase induces the tipping angle. The first sharp increase is reached when electrons have enough energy to cross to the barrier region. The missing transmitted electrons have spin components mostly along $\hat{\mathbf{n}}$  (generated with heavy holes). This increase is then suppressed (plateau region) by the backward transmission from the $\rm{Al_{0.05}Ga_{0.95}As}$ \cite{E_field}. The following drop is due to the transmitted hot electrons generated with light holes  from both sides of the barrier (see Fig.~\ref{fig:scheme}(a) for the differences in spin-momentum correlation of the heavy and light holes cases). Further increase of photon energy suppresses the correlation induced tipping mechanism due to the increased precession rates.

\begin{figure}
   \centering
   \includegraphics[width=8.6cm]{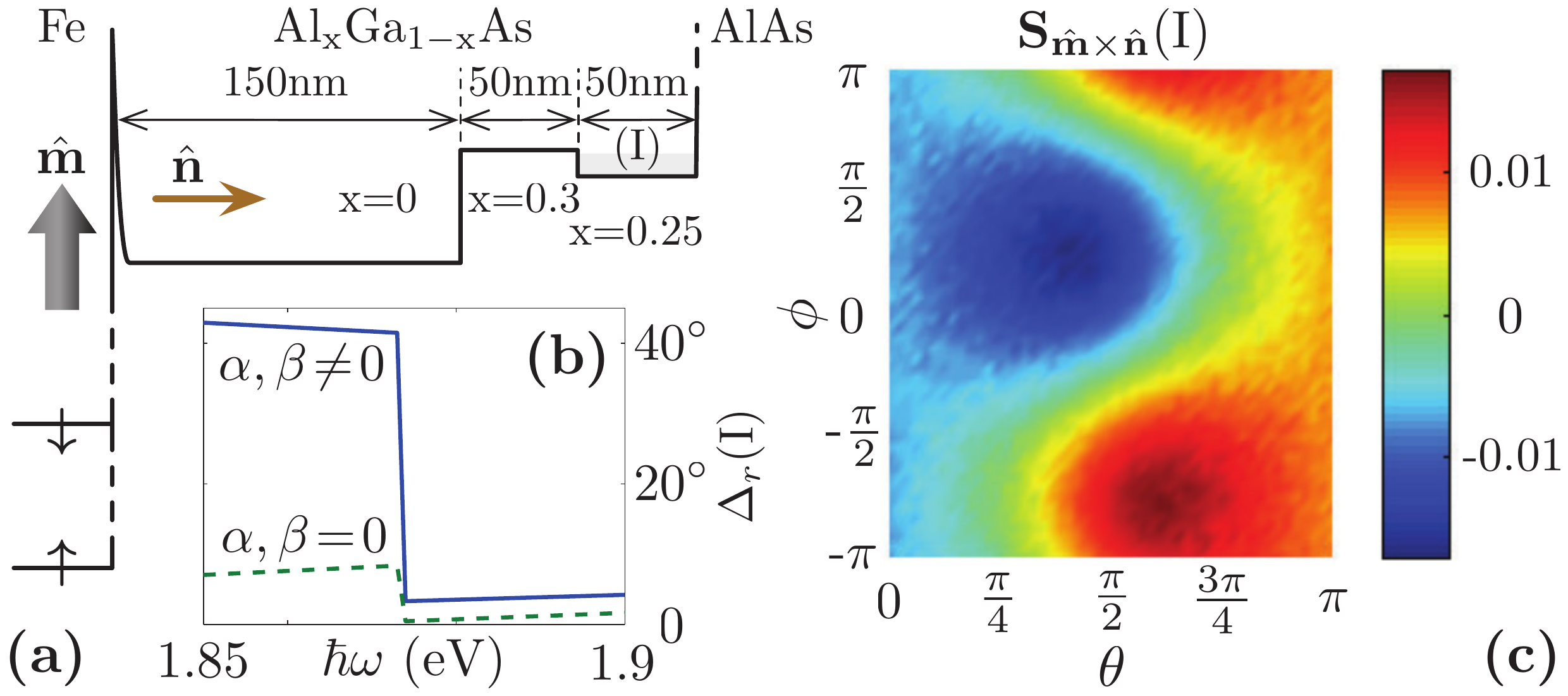}
   \caption{(a) Simulated excitation geometry and heterostructure. $\hat{\mathbf{m}}$, $\hat{\mathbf{n}}$ and $\hat{\mathbf{m}}\times\hat{\mathbf{n}}$ are along crystallographic axes (in-plane magnetization). (b) Tipping angle of the relaxed net-spin vector in region I vs photon energy when the light propagation axis is 45$^\circ$ from $\hat{\mathbf{n}}$. The contribution of alignment and correlation are clearly seen. (c) Relaxed net-spin along $\hat{\mathbf{m}}\times\hat{\mathbf{n}}$ in region I versus light propagation direction (the photon energy is 1.85~eV). The polar ($\theta$) and azimuthal ($\phi$) angles are measured from $\hat{\mathbf{n}}$ and $\hat{\mathbf{m}}\times\hat{\mathbf{n}}$, respectively. } \label{fig:S_im}
\end{figure}

Partial reflection off a ferromagnet provides unique signatures of momentum alignment and spin-momentum correlation that are interwoven with the magnetization direction of the ferromagnet. We perform detailed Monte Carlo simulations for the heterostructure in Fig.~\ref{fig:S_im}(a). It consists of $\rm{Fe/GaAs(150\,nm)/Al_{0.3}Ga_{0.7}As(50\,nm)}$ $\rm{/Al_{0.25}Ga_{0.75}As(50\,nm)/AlAs}$. For certain photon energies reflected hot electrons from the Fe/GaAs interface reach region I ($\rm{Al_{0.25}Ga_{0.75}As}$) only if they do not experience energy relaxation in the GaAs region. In addition, the spins of these electrons do not precess in their short passage since their motion is along a crystallographic axis ($\hat{\mathbf{n}}=\hat{\mathbf{z}}$).  Partial and spin selective reflection across the Fe/GaAs interface is modeled by a 0.5~eV high and 6~nm wide parabolic Schottky barrier \cite{auxiliary}.

Figure~\ref{fig:S_im}(b) shows the tipping angle after energy relaxation in region I as a function of the photon energy. At photon energies just below the band-gap of region I, transmission from the GaAs into region I across the inner barrier is possible only for electrons that are generated with heavy-holes and that are directed along $\hat{\mathbf{n}}$. This is shown in the low energy side of Fig.~\ref{fig:S_im}(b). Since the spin direction of these electrons is parallel to their wavevector, the tipping angle is simply the angle between the optically injected direction and $\hat{\mathbf{n}}$. The step occurs when electrons are photoexcited in region I and the net-spin becomes aligned with the optically injected spin direction (the transmitted electrons from the GaAs region are shadowed by the density of photoexcited electrons in region I). Figure~\ref{fig:S_im}(c) shows the relaxed spin component along $\hat{\mathbf{m}}\times\hat{\mathbf{n}}$ in region I as a function of the light propagation direction for photon energy of 1.85~eV. This relatively small component is an entwined signature of the ferromagnet and the spin-momentum correlation.

The simulated results in Figs.~\ref{fig:Al}~\&~\ref{fig:S_im} are consistent with a simple quantitative analysis in which the precession and correlation induced tipping are treated separately. For both mechanisms, the average spin after relaxation in the non-magnetic structure reads $\mathbf{S}=S_0\left[(1-\delta_a)\hat{\mathbf{p}}-\delta_b(\hat{\mathbf{p}}\cdot\hat{\mathbf{n}})\hat{\mathbf{n}}\right]$. In the optimal case ($\hat{\mathbf{p}}\cdot\hat{\mathbf{n}}=1/\sqrt{2}$) the tipping angle is $\Delta\approx\delta_b/2$ where for the precession induced tipping (setup (i) in Fig.~\ref{fig:Al}(a)) \cite{auxiliary},
\begin{equation}
\Delta_p\approx\frac{\alpha_c^2\tau_{LO}^2}{2\hbar^2E_g}\sum_{n=0}^N \gamma_{p,n}(\hbar\omega-E_g-nE_{LO})^3\,. \label{eq:angle_p}
\end{equation}
$\gamma_{p,n=0}\approx0.06$ and $\gamma_{p,n>0}\approx0.08$ are integration parameters, $E_g$ is the energy gap, and $\alpha_c$ is a measure of the spin-orbit coupling strength in the conduction band \cite{Dyakonov_Perel}. $E_{LO}$ is the longitudinal-optical phonon energy and $\tau_{LO}$ is the associated scattering time. $N=\lfloor(\hbar\omega-E_g)/E_{LO}\rfloor$ denotes the number of phonon emissions prior to relaxation to the bottom of the band. For the correlation induced tipping (setup (ii)) we get,
\begin{equation}
\Delta_c\approx \frac{\beta_0}{2S_0} \gamma_c \left( \vartheta^{3/2}-1.5\vartheta^2+0.1\vartheta^{5/2}+\vartheta^3 \right)\,, \label{eq:angle_c}
\end{equation}
in the low photon energy regime (before the plateau in Fig.~\ref{fig:Al}(e)). $\gamma_c\approx0.4$ is an integration parameter and $\beta_0$ corresponds to the correlation parameter of electrons generated from heavy holes. $\vartheta=[m_{hh}(\hbar\omega-E_g)]/[(m_e+m_{hh})E_1] - 1$ where $m_{hh}$ ($m_e$) is the heavy-hole (electron) effective mass and $E_1$ is the barrier height (see Fig.~\ref{fig:Al}(a)). Using Eqs.~(\ref{eq:angle_p})-(\ref{eq:angle_c}), the spin orbit coupling parameters $\alpha_c$ and $\beta_0$ can be extracted from experiments \cite{auxiliary}. In the magnetic case of Fig.~\ref{fig:S_im}(a) we can also work out the tipping angles since precession effects are small (electrons reach the interface prior to momentum scattering). Here, the integrated surface density of the net-spin vector after magnetic reflection reads,
\begin{eqnarray}
\!\!\!\!\!\!\mathbf{S}_r \!&=&\! n_0\ell S_0 \hat{\mathbf{p}} -n_0\big\{\big(4\lambda_0S_0 + \delta^\beta_0 \big)\hat{\mathbf{p}} - 3(\hat{\mathbf{p}}\cdot\hat{\mathbf{n}})\delta^\beta_0\,\hat{\mathbf{n}} \nonumber \\ &&\!+ \big[  \lambda_1-\lambda_2 + \tfrac{1}{2}\big( 1-3 \big|\hat{\mathbf{e}}\cdot\hat{\mathbf{n}}\big|^2 \big)(\delta^\alpha_1-\delta^\alpha_2) \big]\hat{\mathbf{m}} \nonumber  \\ &&\!- \big(4\lambda_3S_0 + \delta^\beta_3 \big)\hat{\mathbf{m}}\times\hat{\mathbf{p}} + 3(\hat{\mathbf{p}}\cdot\hat{\mathbf{n}}) \delta^\beta_3\, \hat{\mathbf{m}}\times  \hat{\mathbf{n}} \big\}\,. \label{eq:S_general}
\end{eqnarray}
$\ell^{-1}$ is the light absorption coefficient and $n_0$ is the photoexcited electron density. $\lambda_i$ ($\delta_i$) are isotropic and uncorrelated (momentum aligned and spin correlated) integration parameters that depend on the barrier transmission. The total reflection is spin independent and governed by $\lambda_0$ \& $\delta^\beta_0$ terms, the spin selective reflection by ($\lambda_1-\lambda_2$) \& ($\delta^\alpha_1-\delta^\alpha_2$) terms, and the magnetization induced torque by $\lambda_3$ \& $\delta^\beta_3$ terms. The $\alpha$/$\beta$ superscript indicates signatures of alignment/correlation. For detailed expressions of all terms see \cite{auxiliary}. Previous experimental investigations of the ferromagnetic proximity effect \cite{Kawakami_Science01,Epstein_PRB02} and their ensuing theories \cite{Ciuti_PRL02,Bauer_PRL04,Grindev_JETP03} were focused on ferromagnetic signatures while ignoring the alignment and correlation of photoexcited electrons. Ciuti \textit{et al.} have derived a reduced form of Eq.~\!(\ref{eq:S_general}) in which $\delta_i^{\alpha}$=\,$\delta_i^{\beta}$=\,0 \cite{Ciuti_PRL02}. Our analysis shows that in properly designed structures, correlation induced signatures are experimentally resolvable by the $\hat{\mathbf{p}}\cdot\hat{\mathbf{n}}$ amplitude dependence of the non-magnetic ($\hat{\mathbf{n}}$) and magnetic ($\hat{\mathbf{m}}\times  \hat{\mathbf{n}}$) spin components.

In conclusion, reflections off non-magnetic semiconductor heterojunctions and semiconductor/ferromagnet interfaces have been shown to be a powerful tool to study coherent effects of the crystal symmetry and spin-orbit coupling. Tipping the net-spin vector out of the optically injected direction is a measure of these effects. The predicted tipping angles are noticeable and can be probed, for example, via the photoluminescence of energy relaxed free excitons. The tipping angle corresponds to the angle at which the detected circular polarization is maximal. Tunable parameters are the photon energy and light propagation direction. The tipped net-spin vector would ultimately evolve in hyperfine interaction and polarize the nuclear spin system. In this case, ultrafast decaying coherent effects that result from the crystal symmetry and spin-orbit coupling can be inferred by the 10$^{10}$ slower dynamic polarization of the nuclear system.

This work is supported by AFOSR and NSF (Contracts No. FA 9550-09-1-0493 and ECCS-0824075). We deeply thank Dr. Scott Crooker for helpful insights.


\begin{thebibliography}{99}
\bibitem{footnote_Al5Ga95} We choose $\rm{Al_{0.05}Ga_{0.95}As}$ in the right region in order to enhance the net transmission out of the GaAs region. This choice also allows one to distinguish the free exciton luminescence of the GaAs region.
\bibitem{E_field} Application of an external electric field or the use of structures with built-in electric field (e.g., by continous variation of the $\rm{Al}$ content in the barrier region) would enhance the transmission from the GaAs region and it will also suppress the plateau appearance.
\end{thebibliography}
\end{document}